\documentclass[]{elsarticle}
\usepackage{listings}
\usepackage{hyperref}
\usepackage{astrojournals}
\usepackage{booktabs}
\usepackage{multirow}

\journal{Astronomy and Computing}

\biboptions{authoryear}

\begin{document}

\begin{frontmatter}

\title{Minimal Re-computation for Exploratory Data Analysis in Astronomy}

\author{Bojan~Nikolic}
\address{Cavendish Laboratory, University of Cambridge, United Kingdom}
\author{Des~Small and Mark~Kettenis}
\address{Joint Institute for VLBI ERIC, Dwingeloo, The Netherlands}

\begin{abstract}
  We present a technique to automatically minimise the re-computation
  when a data analysis program is iteratively changed, or added to, as
  is often the case in exploratory data analysis in astronomy. A
  typical example is flagging and calibration of demanding or unusual
  observations where visual inspection suggests improvement to the
  processing strategy. The technique is based on memoization and
  referentially transparent tasks. We describe the implementation of
  this technique for the CASA radio astronomy data reduction package.
  We also propose a technique for optimising efficiency of storage of
  memoized intermediate data products using copy-on-write and block
  level de-duplication and measure their practical efficiency.  We
  find the minimal recomputation technique improves the efficiency of
  data analysis while reducing the possibility for user error and
  improving the reproducibility of the final result. It also aids
  exploratory data analysis on batch-schedule cluster computer
  systems.
\end{abstract}


\end{frontmatter}

\section{Introduction}

Notwithstanding the notable successes in automating the reduction and
analysis of data from radio telescopes, the traditional
astronomer-driven data reduction is still common. This typically takes
the form of exploratory, iterative, data reduction where visual
inspection of intermediate or final data products is used to adjust,
or add to, the data processing program. Each adjustment is typically
small and impacts only a subset of the processing as a whole; however
in current systems there is no way automatically re-run only this
subset. Instead the user has the choice to either re-run the whole
program, which can take minutes, hours or even days; or to manually do
the sub-selection of the part of the program which needs to be re-run
and risk introducing errors.

The situation can be easily be improved as we show below, by thin
wrappers around existing data reduction software systems and applying
techniques used in other parts of software engineering. We expect the
benefits will be the greatest during telescope commissioning, during
which the automated pipelines are being developed, and to particularly
demanding observations where careful inspection of flagging,
calibration, CLEANing and uv-weighting may be needed.

Similar workflows are encountered in optical observational astronomy
as well as in general in data analysis. The reason for focusing on
radio-astronomy are very large data-volumes (e.g., a number of radio
telescopes record at a rate of excess of 1 TeraByte/hour) combined
with complex data analysis which is made up out of a number of simpler
tasks. The complexity of analysis in radio astronomy stems from:
\begin{enumerate}
\item Use of \emph{self-calibration} \citep[][]{1984ARA&A..22...97P}:
  iteratively solving for calibration solutions while simultaneously
  improving the estimated image of the sky
\item Iterative flagging of bad data
\item Use of user-supplied explicit constraints during deconvolution
  (`CLEAN boxes').
\end{enumerate}
Self-calibration is illustrated in the sample program shown in
\ref{l:gcexample}.

\section{Objective}

The use case we consider is an astronomer who is repeatedly running a
data reduction processing job with some change to the processing logic
between each run. The primary objective is that, without any
intervention from the astronomer, only the processing steps whose
results could have changed (because either their parameters or their
input data changed) are re-executed between the runs. We assume the
processing logic is captured in a `script' and the above implies
astronomer does not need to edit the script in any way to select steps
which are to be executed -- instead the script as a whole is submitted
for every execution.

The rationale for this objective is to, at the same time, improve the
efficiency of data analysis (both in term of the computing time and
the time of the astronomers doing the analysis) while reducing the
possibility of errors resulting from astronomers manually
sub-selecting parts of the script to be run. By only having a single
script whose logic is incrementally improved we expect reproducibility
and understandability will be improved.

The second objective is that the syntax and semantics of the scripts are
as similar as possible to what astronomers are familiar with. In the
field of radio astronomy at least this means Python, or Python-like
syntax and semantics. The rationale for this is to maximise of the
uptake of this software -- very few would be willing to learn a new
paradigm of computing.

The third objective is to enable modularity of the scripts. The
rationale is that as exploitative analysis becomes more driven by
complete scripts those scripts will become increasingly complex and it
will be desirable to have modularity to make them easier to construct
and to read.

\section{Approach}

\begin{figure}
  \includegraphics[width=\textwidth]{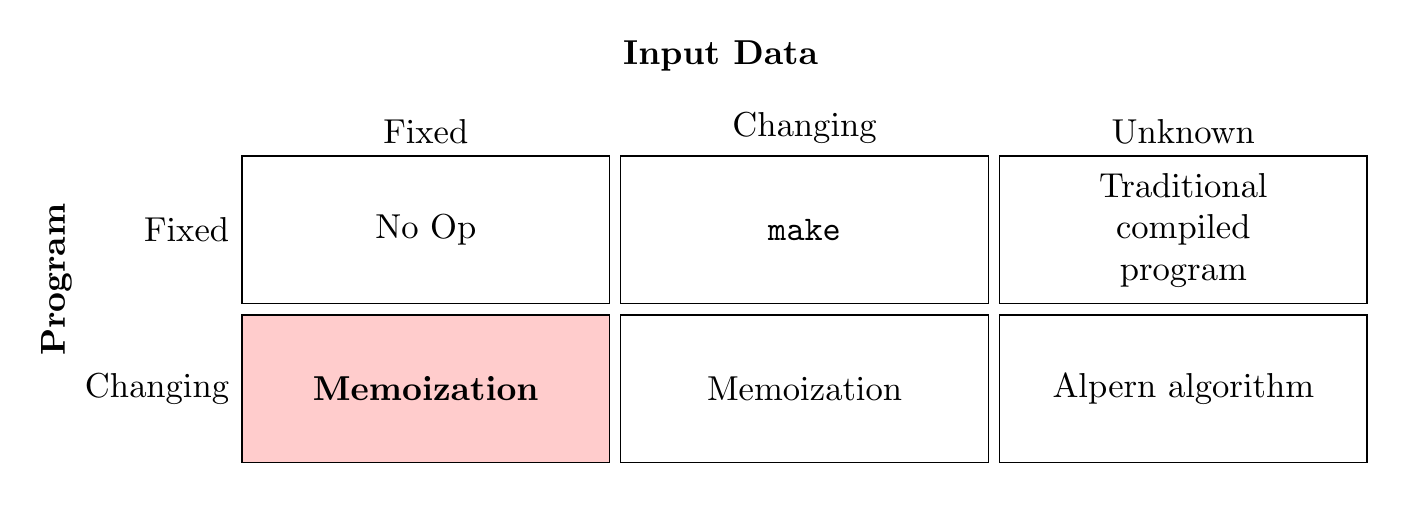}
  \caption{Classification of
  techniques for minimising computation for all combinations of
  changing program and data scenarios. The primary scenario we
  consider here the box shaded red -- input data are
  fixed and the program is changing.}
\label{fig:class}
\end{figure}

Minimal re-computation can be classified based on whether
re-computation happens after a change in the data input or a change in
the program itself. Specifically, the program can be considered either
to be `fixed' or to be `changing' and the data can be considered to be
either 'fixed', 'changing' or `unknown'. The `unknown' category for
data is useful because some program changes can be proven not to
affect the output \emph{regardless} of what data are input. This
classification is illustrated graphically in Fig~\ref{fig:class}.

In the scenario we consider here the input data are known and fixed --
they are the raw data recorded by the telescope -- while the data
processing program is evolving between processing runs. This scenario
can be contrasted with one where the program does not change but some
of the input data changes between runs: the classic example of this is
the {\tt make} program which minimises the re-compilation/re-linking
when a subset of source code files changes.  Another scenario
considered in previous work \citep{Small2015} is where the input data
are unknown, i.e., the minimal set of re-computation needs to be
determined without reference to a particular input data. 

The approach we adopt is to have \emph{referentially transparent
  tasks} at the user level and use \emph{memoization}.  A
referentially transparent task is a task whose call can always be
replaced by its return values.  The important implications for
astronomy are that tasks cannot have observable side effects other
then their return values and that tasks can not modify in-place any of
their input variables. So for example, a task to calibrate a dataset
must return a new calibrated dataset rather than modifying the dataset
it has been passed. The downside of such tasks is that a processing
job will require more disk storage space; we describe below how we
ameliorate this.

Memoization \citep{michie1968memo, abelson1996structure} is the
technique of tabling the results of task invocations against their
inputs. In astronomy this means tabling the results against both the
input astronomical data and any adjustable processing parameters. The
technique is commonly used in functional programming languages or when
a functional sub-set of a language is used.  A well known example is
Packrat parsing \citep{Ford:2002:PPS:583852.581483}. The downside of
memoization is the storage space required for keeping the results,
which we limit with an eviction strategy described below. Another
downside in the general case is the high cost of lookup for extremely
fine grained tasks but this is unlikely to apply to typical
astronomical processing scripts and their tasks.  The approach we
adopt parallels that put forward in some other fields in software
engineering: we list these related works in
Section~\ref{sec:related-work}. As far as we are aware this is the
first time a technique of this type has been applied to scientific
data analysis.

\section{Design \& Implementation}

\begin{figure}
  \includegraphics[width=\textwidth]{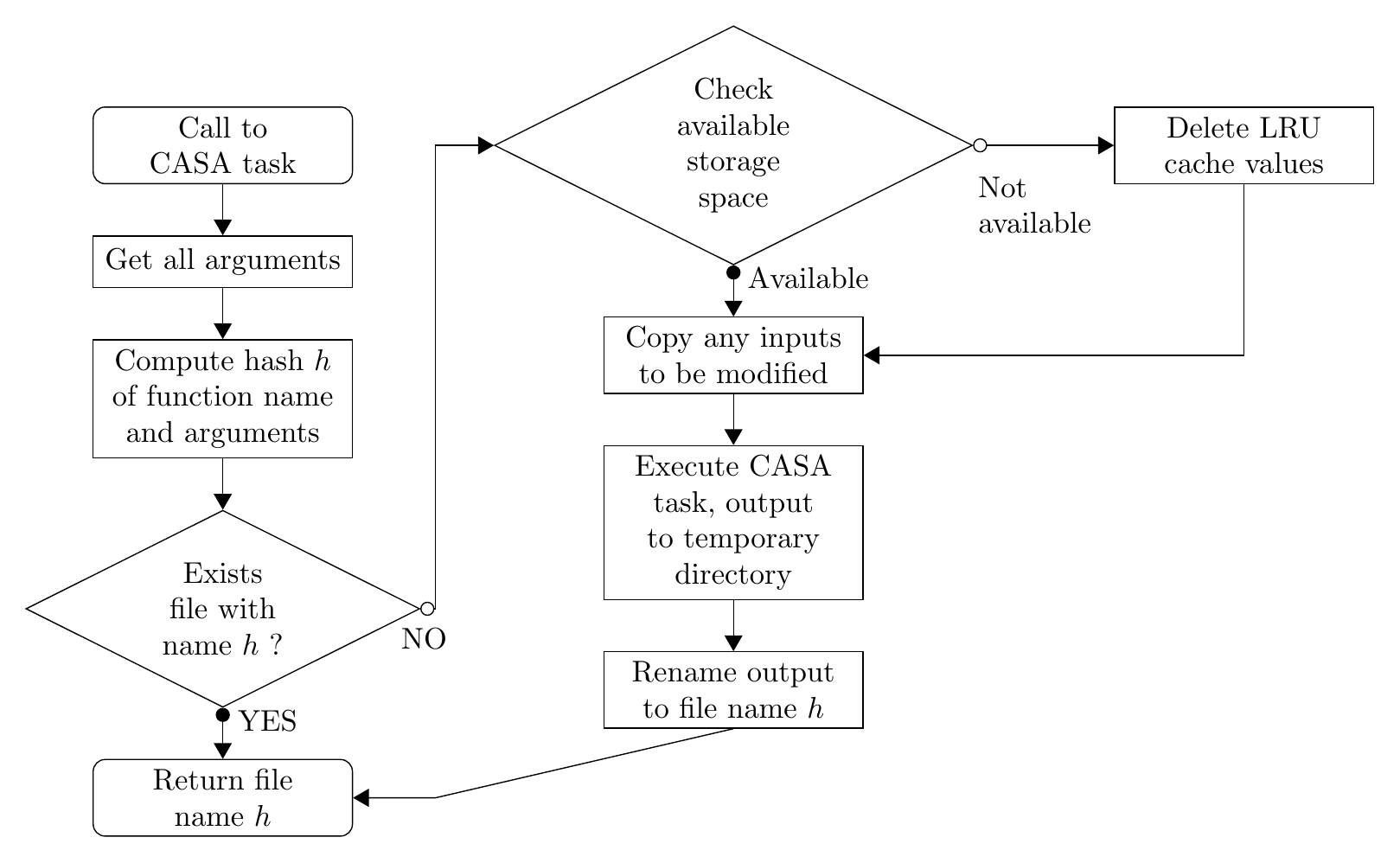}
  \caption{Flow diagram illustrating the algorithm used for minimal
    re-computation applied to CASA workflows as implemented by the
    {\sc recipe\/} system.}
\label{fig:flow}
\end{figure}

The software system we implemented is named Recipe and in this work we
concentrate on using it in cojuction with CASA
\citep{2007ASPC..376..127M}, probably the widest-used software package
in radio astronomy. A similar approach can easily be applied to
ParselTongue \citep{2006ASPC..351..497K} as we showed earlier
\citep{Small2015}. The strategy we adopt is to wrap the existing CASA
interface to make it referentially transparent and to replace the use
of user-defined filenames with automatically generated unique strings
(which the user handles via Python variables rather than
explicitly). We then apply memoization to this new interface, storing
the tabling key as the filename of the value. Values are evicted
according the Least-Recently-Used (LRU) strategy. An outline of the
algorithm is shown in Figure~\ref{fig:flow}.

Users interact with CASA predominately via ``tasks'', a Python-based
interface where input and output data are always files on disk. This
interface is functional in the sense no state is retained in memory
between task invocations. CASA tasks however often create their
outputs by modifying their input datasets, thus invalidating the input
dataset for use in substituting out some previous computation. We make
such tasks suitable for minimal re-computation by copying the input
dataset before the task is invoked and using this copy as the output
value of the task.

Since some tasks only modify a small part of the input dataset, this
copy approach can be inefficient on traditional filesystems. For this
reason we use OpenZFS (on Linux) which supports block-level
de-duplication or BTRFS which supports file-level Copy-On-Write
(COW). With both of those approaches, a copy of a file (i.e., not just
a link to) will not cause any copying of the on-disk blocks until
those specific blocks have been changed for the first time. In the
present case this means that when a task modifies its input dataset,
only the blocks it modifies will actually use disk space. The savings
in disk space due to the use of copy-on-write or de-duplication are
measured in Section~\ref{sec:pract-de-dupl}.

In normal usage, CASA users specify the files which contain inputs and
outputs of tasks directly by their concrete filenames. Multiple
approaches can be used to apply memoization in this situation. One
approach is to hash the input datafiles and use hashes for tabling;
another is to infer the dataflow \citep[see ][ for a
description]{Small2015}. The approach we adopt is to encourage the
analyst to use variables instead of direct filenames, and to ensure
the actual filenames are the string representations of the hash of the
task call that was used to generate their contents. This approach has
two advantages: no hashing of file contents is needed (a potentially
expensive operation) as long as raw input data have stable filenames
and are never modified in-place (the normal situation in
radio-astronomy); and, the resulting code structure is much easier to
modularise since the file variables are lexically scoped in the same
way as the other variables. By contrast, direct use of filenames is
equivalent to a single global namespace for file variables.

Memoized values are checked for eviction before every call into a
task. The trigger is the free disk space on the filesystem used to
store the values, i.e., values will be evicted if the free space is
less than a certain critical value. The criterion used is the
last-access time stamp of the files. This implements the
least-recently-used cache eviction scheme. When a task checks for the
existence of a memoized value, it updates the last-access time even if
the value is not read (i.e., the descendant task has also been
memoized) -- this is done in order to reduce the chance of early
eviction of roots of long, deep dataflows.

Complete information, source code and links to the Git revision
control repository for the full implementation is available at
\url{http://www.mrao.cam.ac.uk/~bn204/sw/recipe}. Introspection
facilities of Python and regularity of the CASA task interface means
that only around 150 lines of Python are required for the
implementation.

A reduced example script using this implementation is shown in
Listing\ref{l:gcexample}. This script implements iterative
self-calibration on the Galactic Centre for HERA as described by
\cite{NikolicURSI20178105387}.  The only differences to typical
CASA/Python scripts are:
\begin{enumerate}
\item Filenames are never specified explicitly by the analyst except
  for the input data (which are assumed not to change)
\item Every wrapped CASA task returns a single value which is the
  filesystem path containing the output of the task
\end{enumerate}
We have found analysts familiar with CASA are able to use the minimal
re-computation framework with little training.

In Table~\ref{tab:timing} we present a comparison the run time
performance of our minimal re-computation software versus plain
CASA. The workload used is a typical self-calibration workload,
consisting of a slightly extended version of the script in
Listing\ref{l:gcexample}. It can be noted that:
\begin{enumerate}
\item On the initial run, performance of Recipe/CASA is close to, but
  slightly better than plain CASA. This is because the automatic
  re-use of some intermediate data product in the self-calibration
  loop outweighs the time for some additional copying of data that
  minimal recomputation requires.
\item On subsequent runs of the same script the performance of
  Recipe/CASA is extremely high as only cache queries corresponding to
  checking of existence of a file need to be made. As there is no need
  to list the whole cache directory, the performance is not expected
  to degrade appreciably with increasing cache size. 
\end{enumerate}

\section{Practical de-duplication and copy-on-write efficiency}
\label{sec:pract-de-dupl}

We have measured the efficiency (with respect to the volume of storage
required) of both the de-duplication and copy-on-write approaches. The
measurement was made with a slightly expanded version of program in
Listing~\ref{l:gcexample} with input data consisting of a single HERA
observation with 40 antennas. The size of the input dataset is 1.8\,GB
and size of the complete repository for intermediate data products is
41\,GB.  All tests were done with CASA version 5.1 using standard
Measurement Set V2 files on GNU/Linux Ubuntu 17.04 Kernel 4.10 with
both ZFS (for de-duplication) and BTRFS (for copy-on-write).

When using ZFS block-level de-duplication the allocated space on disk
is 6.5\,GB and the de-duplication factor $6.2\times$ (both as reported
by the {\tt zpool list} command).  Some copying of data is unavoidable
in normal manual CASA processing, this shows that with de-duplication
is highly efficient for this problem, and that the repository is only
perhaps two times larger than would be necessary at the minimum. ZFS
de-duplication however requires a significant reserve of RAM to
function efficiently as it needs to maintain a table with the
check-sums of all of the blocks.

In the same test environment we measured the efficiency of the
copy-on-write mechanism provided by BTRFS. For this purposes all
copies were done with {\tt cp --reflink=auto} flag. We find disk usage
is 8.4\,GB, a saving of $4.9\times$ compared to no
copy-on-write. Therefore, the copy-on-write mechanism is less
efficient than de-duplication, but still very good. The advantage of
copy-on-write compared to de-duplication is that it does not impose
additional RAM memory requirements. In most general purpose scenarios
we expect the copy-on-write approach would give sufficient storage
efficiency without additional hardware requirements.

\section{Minimal re-computation in a batch-scheduled cluster environment}

Analysis of data from large radio telescopes (e.g., ALMA, JVLA, LOFAR)
is often done on shared-filesystem computer clusters. Such clusters
are often batch-scheduled to facilitate efficiency, making interactive
use impossible.  Exploratory data analysis must then be done by
submitting multiple data analysis scripts for processing which creates
issues of many unnecessary re-computations and keeping track of the
results of each script.

The proposed technique of minimal re-computation and the
implementation we have created work on a multi-user shared-filesystems
cluster without need for modification and with high efficiency. The
key feature is the use of the filesystem as the only store of
information, which being a shared resource on the cluster means that
all computers see a synchronised version of the data. In particular,
both multiple jobs and multiple users can re-use each other's
previously computed values regardless on which computer on the cluster
they get scheduled on.

We find use of minimal recomputation significantly improves the
work-flow of exploratory data analysis on a batch-scheduled cluster:
\begin{itemize}
\item It is efficient to submit initial parts of the script while
  later parts are being developed: the results of initial evaluation
  can then be available when the full script is submitted
\item There is no need to constrain the scheduling of many similar
  scripts: they can be executed in any order and on any machine and
  will re-use each others intermediate values as available
\item The enforced referential-transparency of tasks usefully  aids
  modularisation which is more important when interactive use is not
  possible.
\end{itemize}

\begin{table}
  \begin{centering}
    
\begin{tabular}{@{}llcc@{}} \toprule
\multicolumn{2}{c}{Configuration} \\ \cmidrule(r){1-2}
  \multirow{2}{*}{Environment} & \multirow{2}{*}{Software} &  1$^{\rm st}$ Execution & Repeated Execution\\
   & & Time& Time   \\ \midrule
\multirow{2}{*}{Workstation} & CASA & 10m30s & 10m29s\\ 
                               & Recipe/CASA & 9m59s & 0.1s\\
\multirow{2}{*}{Cluster} & CASA & 16m37s & 16m9s\\ 
 & Recipe/CASA & 19m37s & 0.1s\\  
 \bottomrule
\end{tabular}
\caption{Results of measurement of execution time. The `Workstation'
  environment is a Dell T7810 Single Node Dual Socket Xeon E5-2680 v3
  with 64 GB ECC RAM with 4x Samsung SM961 512GB NVMe Storage
  organised as a striped ZFS filesystem. The `Cluster' environment
  refers to University of Cambridge's `CSD3' cluster with Intel
  Skylake processors and a shared Lustre filesystem connected to
  OmniPath fabric. `Recipe/CASA' refers to our minimal re-computation
  software driving CASA, as described in text. CASA version 5.3 for
  RHEL 7 was used. The script which was timed is a slightly extended
  version of \ref{l:gcexample} representing a realistic
  self-calibration processing in radio astronomy. All times are
  wall-clock times.}
\label{tab:timing}
\end{centering}
\end{table}

Measurement of run-time performance in the cluster environment is
presented in Table~\ref{tab:timing} (together with single-node
performance as described above). We note that overall execution time
slower in the cluster. Experience suggests this primarily due to
overheads of access to the data over the shared Lustre filesystem. We
also note Recipe/CASA in the cluster environment is now slower than
plain CASA: this is because the slower file-system access increases
the overhead due to copying of data inherent in the minimal
re-processing when some operations modify their input
data. Nevertheless, the primary advantage of very fast repeated
execution is clearly demonstrated in the cluster environment.

\section{Related work}
\label{sec:related-work}

The most generally familiar related system is {\tt
  ccache}\footnote{\url{https://ccache.samba.org/}}. It minimises
recompilation when both the source code and the instructions for
compilation may be changing (e.g., due to edits to {\tt Makefile}s or
experimenting with different compilation switches). In this scenario
the equivalent of our data analysis program are the {\tt Makefile}s
while the equivalent to our astronomical data is the program source
code. {\tt Ccache} is specialised for compilation only; also, in
contrast to our system, it always reads the full input data (and also
does pre-processing on them) before computing the hashes.

The most direct inspiration was the Ciel dataflow execution engine
\citep{Murray:2011:CUE:1972457.1972470}.  Another inspiration was the
NiX \citep{DolstraNIX} system, which like {\tt ccache} works in the
domain of software building, but applies the minimal re-computation
methodology to building a whole Linux distribution rather than a
single program. From NiX we take the idea of using the filesystem as a
database keyed by the hash of the processing requested from each task.

A related system is Nectar described by \cite{Gunda2010NectarAM}. It
too uses the idea of a cache of intermediate results (or
sub-expression values) to speed up program execution when the program
is being developed, or shared by multiple users. It includes a
sophisticated rewriting engine rather than simple memoization. Nectar
is specialised on the LINQ data processing language working in the CLR
environment, so would be difficult to integrate with software such as
CASA. A similar related system is IncPy presented by \cite{Guo2011},
This system is also based on memoization, however it requires a
modification to the Python interpreter in order to dynamically analyse
dependencies of Python functions.  It could not therefore be applied
to CASA (which comes with own Python interpreter and accesses files
from its C++ code). The system presented here additionally replaces
all user-created files with unique values based on memoization hashes;
this allows re-use of values across many previous generation of the
analysis script.

\section{Conclusions}

We present an approach to minimal recomputation and a concrete
implementation for the CASA radio astronomy data analysis system. The
implementation can be adapted to other task-based astronomy data
analysis applications which use the filesystem for storage of all
input and output data. When used in combination with a copy-on-write
filesystem, the technique has modest additional storage requirements
while providing significant benefits in reproducability and efficiency
of the data analysis. The benefits are amplified in batch-scheduled
computer clusters where interactive use is difficult.

The approach presented here has implications on the design and
implementation of data models used in science: models structured so
that changes or additions to data change as few on-disk block as
possible will provide highest efficiency in terms of storage
requirements.

The technique here also significant bearing on `notebook-style'
interaction with programs, as first popularised by Wolfram Research's
\emph{Mathematica}, and now often used in the field of scientific data
analysis through the Jupyter\footnote{\url{http://jupyter.org/}}
  system. In such systems, when an entry in a notebook is changed, the
  user usually has the options to either re-evaluate the whole
  notebook or try to singly evaluate all statements which my be
  affected by the first change. Minimal re-computation resolves this
  problem (subject to sufficient storage space) by allowing the user
  to always select re-evaluation the whole notebook while
  automatically only the statements which could have changed are
  actually re-evaluated.

\section*{Acknowledgements}

We are pleased to acknowledge the support of EC FP7 RadioNet3/HILADO
project (Grant Agreement no.  283393) and the EC ASTERICS Project
(Grant Agreement no. 653477). We thank A. Madhavapeddy who via Peter
Braam pointed us toward the memoization approach. We also thank Peter
Braam for careful reading of an initial version of the manuscript. We
thank the anonymous referee for helpful comments that improved the
article.

\section*{References}

\bibliography{minimal}

\appendix

\section{Example Recipe program}

\begin{lstlisting}[frame=trBL,language=Python,basicstyle=\ttfamily\footnotesize,label=l:gcexample,caption={Self-calibration
  data reduction based on script for imaging data from the Hydrogen
  Epoch of Reionization Experiment (HERA). Multiple iterations of the
  self-call loop are made with different strategies (number of CLEAN
  iterations, gain versus bandpass calibration). }]
# Input, unchaging files
ingc="GCmodl.cl"
invis="exampleinputvis.ms"

# Insert model for first calibration
f1=ft(invis, complist=ingc, usescratch=True)

# Calculate delay calibration
K=gaincal(f1,  gaintype='K')

# Calculate antenna based-gain errors
G=gaincal(f1, gaintable=[K],   gaintype='G')

# Apply the initial solutions
f2=applycal(f1,  gaintable=[K, G])

# Outputs can be assigned directly to other variables 
f3=f2

# First clean, to setup a very simple model for the GC
ooorig=clean(vis=f3, niter=50)

# Example of a clean Module
def iterselfcal(visin, img, niter,  dobandpass=False):
    """Make one iteration of the self cal loop
    :param img: Image to use as the model
    :param niter: Number of minor cycle iterations
    :param dobandpass: Do a bandpass calibration ?
    """

    # Now back to the original vis for the calibration. NB. use the
    # *model* output from previous here
    f33=ft(visin,
           model=os.path.join(img, "img.model"),
           usescratch=True)
    ctl=[]
    K2=gaincal(f33, gaintype='K')
    ctl.append(K2)

    G2=gaincal(f33, gaintype='G', gaintable=[K2])
    ctl.append(G2)

    if dobandpass:
        B2=bandpass(f33, bandtype="B", gaintable=ctl)
        ctl.append(B2)

    f33=applycal(f33, gaintable=ctl)        
    # Reclean
    oo=clean(vis=f33, niter=niter)
    return oo, ctl

# Incrementally CLEAN and re-solve for solutions, first with and then
# without bandpass
oo1 ,GG1=iterselfcal(f1, ooorig, 500)
oo11,GG2=iterselfcal(f1, oo1, 5000)
oo2 ,GG3=iterselfcal(f1, oo1, 500, dobandpass=True)
oo3 ,GG4=iterselfcal(f1, oo2, 5000, dobandpass=True)
\end{lstlisting}

\end{document}